\documentclass{IEEEtran}
\usepackage{graphicx}
\usepackage{cite}
\usepackage{amsmath,amssymb,mathtools,bm,etoolbox,fancyhdr}
\usepackage{array}
\usepackage[caption=false,font=footnotesize]{subfig}
\usepackage{fixltx2e}
\usepackage{stfloats}
\usepackage{graphicx}
\usepackage{float}
\usepackage{url}
\usepackage{bm}
\usepackage{multirow}
\usepackage{xcolor}
\renewcommand\IEEEkeywordsname{Index Terms}
\IEEEoverridecommandlockouts
\begin{document}
\title{SparseCast: Hybrid Digital-Analog Wireless Image Transmission Exploiting Frequency Domain Sparsity \thanks{This work received support from the European Research Council (ERC) through Starting Grant BEACON (agreement 677854).}\thanks{T.-Y. Tung is with the University of Southern California Viterbi School of Engineering, Los Angeles, CA 90089 USA (email: tzeyangt@usc.edu).}\thanks{D. G\"und\"uz is with the Department of Electrical and Electronics Engineering, Imperial College London, London SW7 2AZ, U.K. (email: d.gunduz@imperial.ac.uk).}}
\author{Tze-Yang Tung and Deniz G\"und\"uz\vspace{-0.5cm}}
\maketitle

\begin{abstract}
A hybrid digital-analog wireless image transmission scheme, called SparseCast, is introduced, which provides graceful degradation with channel quality. SparseCast achieves improved end-to-end reconstruction quality while reducing the bandwidth requirement by exploiting frequency domain sparsity through compressed sensing. The proposed algorithm produces a linear relationship between the channel signal-to-noise ratio (CSNR) and peak signal-to-noise ratio (PSNR), without requiring the channel state knowledge at the transmitter. This is particularly attractive when transmitting to multiple receivers or over unknown time-varying channels, as the receiver PSNR depends on the experienced channel quality, and is not bottlenecked by the worst channel. SparseCast is benchmarked against two alternative algorithms: SoftCast and BCS-SPL. Our findings show that the proposed algorithm outperforms SoftCast by approximately 3.5 dB and BCS-SPL by 15.2 dB. 
\end{abstract}


\section{Introduction}
\label{sec:introduction}

Conventional wireless image/video transmission systems consist of two components: a source encoder for compression, and a channel encoder that introduces redundancy against noise and interference. This separate design is without loss of optimality according to Shannon's separation theorem, and has dominated practical implementations. Although the optimality breaks down for multi-user systems or time-varying channels, there has been a steady move towards fully digital and separate architectures (e.g., digital TV/radio) thanks to the modularity and flexibility it provides. However, many emerging applications from tactile Internet to autonomous vehicles require wireless transmission of image/video files under extreme latency, energy and complexity constraints, which preclude the use of advanced compression and channel coding techniques. 

A surprising result in \cite{goblick_theoretical_1965} shows that, when transmitting independent Gaussian samples over a Gaussian channel, with one sample per channel use on average, uncoded transmission, where each sample is simply scaled and transmitted, meets the theoretical Shannon bound. With digital transmission, the same performance can only be achieved by vector-quantising an arbitrarily long sequence of source samples, followed by a capacity achieving channel code. Benefits of analog transmission has since been shown in various settings \cite{AguerriGunduz:IT:16}, \cite{LapidothTinguely:IT:10}. We highlight that, analog transmission here does not refer to traditional analog modulation techniques, i.e., amplitude or frequency modulation. Instead, it refers to a transmission scheme, in which both the source and channel encoder/decoder employ sampling; however, the samples are allowed to take continuous values, rather than being limited to a discrete set of quantised values or constellation points. 

Motivated by the theoretical properties of uncoded transmission, a practical joint source-channel coding (JSCC) scheme, called SoftCast, was proposed in \cite{jakubczak_softcast:_2010}. SoftCast applies a discrete cosine transform (DCT) on the image, and transmits the DCT coefficients directly over the channel using a dense constellation. Compression is obtained by discarding blocks of DCT coefficients whose energy is below a threshold. Index of the discarded blocks is sent as meta-data to the receiver for reconstruction. Since the encoder is linear and the coefficients are corrupted by additive noise directly, the resultant video peak signal-to-noise ratio (PSNR) is linearly related to the channel signal-to-noise ratio (CSNR), solving the \textit{cliff effect} problem encountered in separate source and channel coding. 

This letter aims to reduce the bandwidth usage in uncoded image/video transmission by utilising a novel grouping of DCT coefficients, and by incorporating compressed sensing (CS) and sparse signal recovery. CS theory demonstrates that a system of underdetermined equations can be solved with high probability if the solution is sparse \cite{donoho_compressed_2006}. This implies that if an image or a video frame can be transformed into a sparse domain, even if we send a few linear combinations of the pixel values to the receiver, it can still recover the original frame. CS has been previously used for wireless video transmission in \cite{li_video_2011} \cite{schenkel_compressed_2010}, where the $l_1$ approximation for recovery is considered. While this allows the receiver to employ convex optimisation, it may still be computationally complex for video streaming applications with strict delay constraints. Iterative algorithms have been developed to approximate the solution faster at the cost of greater error. One such algorithm, used in \cite{yin_compressive_2016} and \cite{wang_wireless_2014}, called block CS-smooth projected Landweber (BCS-SPL) \cite{mun_block_2009}, achieves reconstruction through iterative thresholding. This algorithm applies CS on the pixels directly and requires no meta-data as long as the measurement matrix is agreed a priori between the transmitter and receiver. It is important to note here that in \cite{yin_compressive_2016}, optimal power allocation is not considered and an additional image processing technique was employed following BCS-SPL to improve the output image quality. 

In this letter, after applying 2D-DCT on image blocks and thresholding, a novel grouping of the coefficients is applied, where coefficients of the same frequency component are grouped into vectors. We then multiply each vector with a pseudo-random measurement matrix whose size depends on the sparsity level of the corresponding vector. Finally, a scaling factor is applied to the results of this multiplication, which corresponds to power allocation across different frequency components. The receiver employs a combination of approximate message passing (AMP) \cite{donoho_information-theoretically_2013} and minimum mean squared error (MMSE) estimation. AMP is a low-complexity iterative thresholding algorithm for CS recovery, which does not need to know the exact positions of the nonzero elements in the sparse vector. It converges exponentially and does not assume any prior distribution on the data, incurring a minimal computation overhead.

The benefits of SparseCast can be summarised as follows: i) Instead of removing blocks as in SoftCast, it uses thresholding on individual coefficients and CS to reduce the channel bandwidth; ii) By grouping coefficients of different blocks according to their frequencies, it better exploits the sparsity of higher frequency components; iii) Unlike other CS-based image/video transmission schemes, it employs power allocation according to empirical variances, which significantly improves the reconstruction quality. 

\section{Proposed Algorithm}
\label{sec:proposed algorithm}

\subsection{Encoder}
\label{subsec:encoder}

\begin{figure}
	\includegraphics[width=\linewidth]{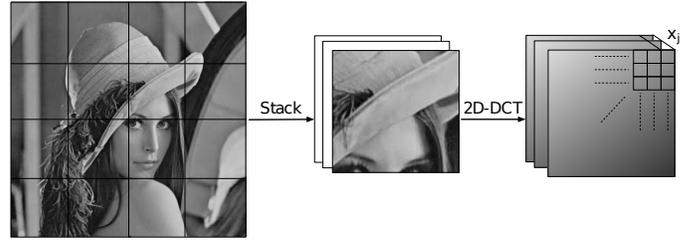}
	\caption{Encoder divides the frame into blocks of $\sqrt{b}\times\sqrt{b}$ before computing the 2D-DCT coefficients, and forms vectors $\mathbf{x}_j$ by extracting the DCT coefficients corresponding to the same pair of indices across all blocks.}
	\label{fig:DCT_grouping}
\end{figure}

The image is divided into $N$ non-overlapping blocks of size $\sqrt{b}\times\sqrt{b}$, and 2D-DCT is applied to each block before being stacked on top of each other to form a 3D matrix of size $\sqrt{b}\times\sqrt{b}\times N$. Vectors $\mathbf{x}_j\in\mathbb{R}^N,~j=1,...,b$ along the third dimension are formed corresponding to each pair of row and column indices, as illustrated in Figure \ref{fig:DCT_grouping}, before being ``sparsified" by setting values that are smaller than the sparsity threshold $\tau_K$ to zero. This means that the same frequency components of all the DCT blocks are stacked into the same vector. Most natural images have DCT energy focused at low frequency components, therefore high frequency components will likely be set to zero and have minimal effect on the image quality.

The \textit{sparsity level} of vector $\mathbf{x}_j$ refers to the number of its non-zero entries, and is denoted by $k_j$, i.e., $k_j \triangleq ||\mathbf{x}_j||_0$. The empirical mean of each $\mathbf{x}_j$ vector is first subtracted from all its entries to obtain a zero-mean vector, such that the empirical variance of each vector is equivalent to its power, before being multiplied by a pseudo-random orthonormal measurement matrix $\bm{\Phi}_j\in\mathbb{R}^{\mu_j\times N}$, where $\mathbf{\tilde{y}}_j \triangleq \bm{\Phi}_j (\mathbf{x}_j-c_j)$ and $c_j$ is the empirical mean of vector $\mathbf{x}_j$. We wish to maintain $k_j\leq\mu_j\ll N$, where $\mu_j$ is the number of measurements sent for vector $\mathbf{x}_j$, which provides a trade-off between the accuracy and the channel bandwidth. Note that, to distinguish the received signals for each of the vectors, the decoder needs to know $\mu_j$'s, which increases the meta-data size. To reduce the meta-data, we choose $\mu_j$'s from a set of $S$ predefined measurement levels. Thus, for each vector, $\mu_j$ is chosen as the closest value among the set of predefined measurement levels that is higher than the desired number of measurements. The total amount of meta-data for $\mu_j$'s to be transmitted is then given by $b\cdot\log_2S$ bits. Moreover, in the case where $\mu_j=N$, we set $\bm{\Phi}_j = \mathbf{I}$ and $\mathbf{x}_j$ is not sparsified as to not lose performance when there is insufficient sparsity to exploit.

Finally, we allocate the transmit power among different frequency components according to their empirical variances. We would like to transmit the elements of these vectors over the channel by simple scaling. The optimal scaling coefficients to transmit independent Gaussian samples, derived in \cite{lee_optimal_1976}, is presented in Lemma 1 for completeness.

\emph{Lemma 1}: Given $L$ data vectors $\mathbf{x}_1,...,\mathbf{x}_L$, each consisting of $m_j=|\mathbf{x}_j|$ samples from a zero mean Gaussian distribution with variance $\lambda_j$, for $j=1,...,L$, the linear encoder that minimises the MMSE in the presence of additive white Gaussian noise (AWGN) scales the $j$-th vector by $g_j$, where
\begin{align}
	g_j&=\lambda_j^{-1/4}\bigg(\sqrt{\frac{\sum_{j}m_j}{\sum_j m_j\sqrt{\lambda_j}}}\bigg),\quad j=1,\ldots,L.
\end{align}
In our implementation, we have $m_j = \mu_j$, the number of measurements transmitted for $\mathbf{x}_j$, $j=1,\ldots,b$. 

\begin{figure}
\centering
	\includegraphics[width=0.8\linewidth]{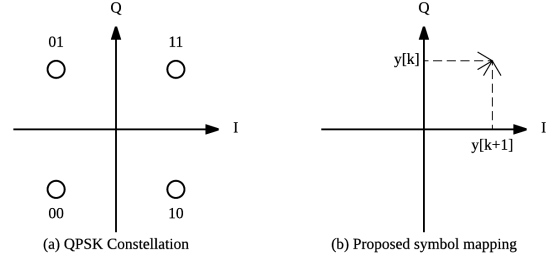}
	\caption{Unlike traditional modulation schemes, such as QPSK, SparseCast does not have a fixed set of constellation points.}
	\label{fig:PHY_mapping}
	\vspace{-0.4cm}
\end{figure}

Channel symbols are formed by pairing consecutive elements of vector $\mathbf{y}_j=g_j \mathbf{\tilde{y}}_j$ as the in-phase and quadrature (I/Q) components of a complex symbol (Figure \ref{fig:PHY_mapping}). This differs from traditional systems that map a bit sequence to a predefined set of discrete constellation points, and allows the channel noise to directly corrupt the coefficients; and hence, preserves the linearity of the scheme (apart from the thresholding, which serves for compression). The generated symbols along with pilot symbols for channel estimation are sent over the channel.

Meta-data consisting of the empirical mean and variance values as well as the number of measurements $\mu_j$ for each $\mathbf{x}_j$ are sent separately to the receiver. For the proposed technique to work, the meta-data must be received without error; therefore, we employ BPSK modulation with 1/2 rate convolutional code for a strong protection against channel errors.

\subsection{Decoder}
\label{subsec:decoder}

The received vector is $\mathbf{\hat{y}}_j=\mathbf{y}_j+\mathbf{n}_j$, where $\mathbf{n}_j$ is the AWGN term with power $\frac{1}{2}\sigma^2_{\mathbf{n}},\hspace{0.5mm}\forall j$; the noise power is assumed to be the same for all transmissions. The decoder employs AMP to decode the sparse $\mathbf{x}_j$ vector if $\mu_j<N$. When $\mu_j=N$, MMSE estimator is used, as the AMP algorithm performs poorly for non-sparse vectors.

The MMSE estimate of a non-sparse $\mathbf{x}_j$ vector is given by $\mathbf{\hat{x}}_j = \frac{g_j\lambda_j}{g_j^2\lambda_j+\sigma^2_{\mathbf{n}}/2}\mathbf{\hat{y}}_j+c_j$, where $c_j$ and $\lambda_j$ (i.e. empirical mean and variance) of the vector $\mathbf{x}_j$ is obtained from the meta-data. The 2D-DCT coefficients of the image are then reconstructed from the $\mathbf{\hat{x}}_j$ vectors before inverse 2D-DCT is performed to obtain the original pixels. 

Although AMP increases the complexity of decoding compared to the linear decoding in SoftCast, its per iteration complexity is dominated by a matrix-vector multiplication operation with a worst-case complexity of $O(\mu_j N)$ \cite{donoho_information-theoretically_2013}, and the required number of iterations is typically on the order of tens. As we will see in the next section, this slightly increased decoding complexity can be justified with the increased performance.

\vspace{-0.2cm}

\section{Results}
\label{sec:results}

We compare SparseCast with sparse recovery using BCS-SPL as described in \cite{yin_compressive_2016}, SoftCast and standard digital encoding by comparing the achieved PSNR across a range of CSNR values.  A smooth and linear relationship between PSNR and CSNR is desirable particularly when transmitting to multiple receivers or over a time-varying unknown channel, while a higher PSNR for given CSNR and channel bandwidth constraints is indicative of better bandwidth utilisation efficiency. The algorithms are first simulated in MATLAB and then implemented using USRP NI2900 and LabView Communications Design Suite 2.0 for real world testing.

\begin{figure}[h!]
	\centering
	\includegraphics[width=\linewidth]{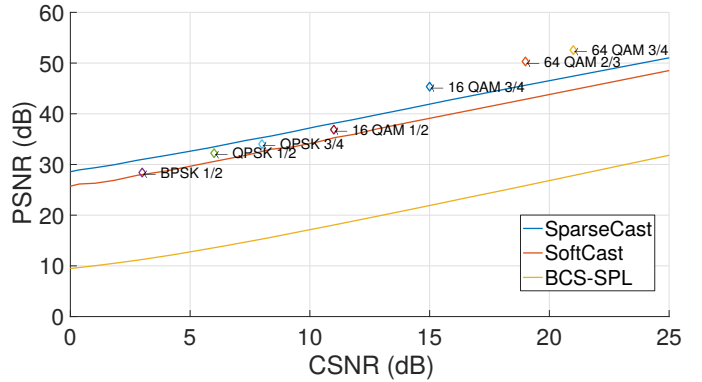}
	\caption{PSNR vs. CSNR. The channel symbol length is $131,000$ symbols. The single points correspond to the PSNR values at the CSNR threshold of each modulation scheme from the 802.11a standard.}
	\label{fig:digital_analog_sim} 
	\vspace{-0.2cm}
\end{figure}

Figure \ref{fig:digital_analog_sim} shows a linear relationship between CSNR and PSNR for the considered uncoded transmission schemes, indicating they can achieve strong multicasting performance, and are less sensitive to inaccuracies or the lack of channel state information at the transmitter. SparseCast is approximately 3.5 dB and 15.2 dB better in PSNR than SoftCast and BCS-SPL, respectively, showing the superiority of SparseCast in terms of its ability to adapt to varying CSNR, as well as bandwidth efficiency over a wide range of channel conditions. 

Advantages of uncoded transmission schemes are clear when compared with digital transmission with JPEG compression followed by conventional constellations and codes from the 802.11a standard. The points on Figure \ref{fig:digital_analog_sim} correspond to the achieved PSNR values at the corresponding CSNR threshold. The CSNR threshold for each constellation and convolutional code rate pair with 10\% packet loss rate are given in Table \ref{table:SNR_th_non_code}. Different constellations and code rates have different compression requirements. For example, BPSK ($1~\mbox{bit/symbol}$) with 1/2 rate (input bits/output bits) convolutional code and 131,000 available channel symbols implies the source image size must be compressed under $1~ \mbox{bit/symbol}\times 131000~\mbox{symbols} \times 0.5~\mbox{rate}=65,500~\mbox{bits}$. In contrast, JSCC schemes do not suffer from the cliff effect and can smoothly adjust the output PSNR with respect to CSNR. SparseCast and SoftCast both follow and even sometimes surpass the envelope formed by the different digital modulation schemes, especially at low CSNR values, suggesting their superiority for wireless image transmission while avoiding the cliff effect.

\begin{table}
\caption{CSNR thresholds for uncoded/coded constellations in 802.11a}
\label{table:SNR_th_non_code}
\begin{tabular}{|c|c|c|c|c|c|}
\hline
\multirow{2}{*}{Constellation} & \multirow{2}{*}{Code rate} & \multicolumn{4}{c|}{CSNR threshold (dB)} \\ \cline{3-6} 
        &             & Uncoded &  Coded 1/2  &  Coded 2/3  &  Coded 3/4 \\ \hline
BPSK    & 1/2 or 3/4  &  8      &  3          &  -          &  5         \\ \hline
QPSK    & 1/2 or 3/4  &  11     &  6          &  -          &  8         \\ \hline
16-QAM  & 1/2 or 3/4  &  18     &  11         &  -          &  15        \\ \hline
64-QAM  & 2/3 or 3/4  &  24     &  -          &  19         &  21        \\ \hline
\end{tabular}
\end{table}

\begin{figure} 
    \centering
  \subfloat[High bandwidth usage ($131,000$ channel symbols).]{%
  \label{subfig:131k_USRP_lenna}
       \includegraphics[width=\linewidth]{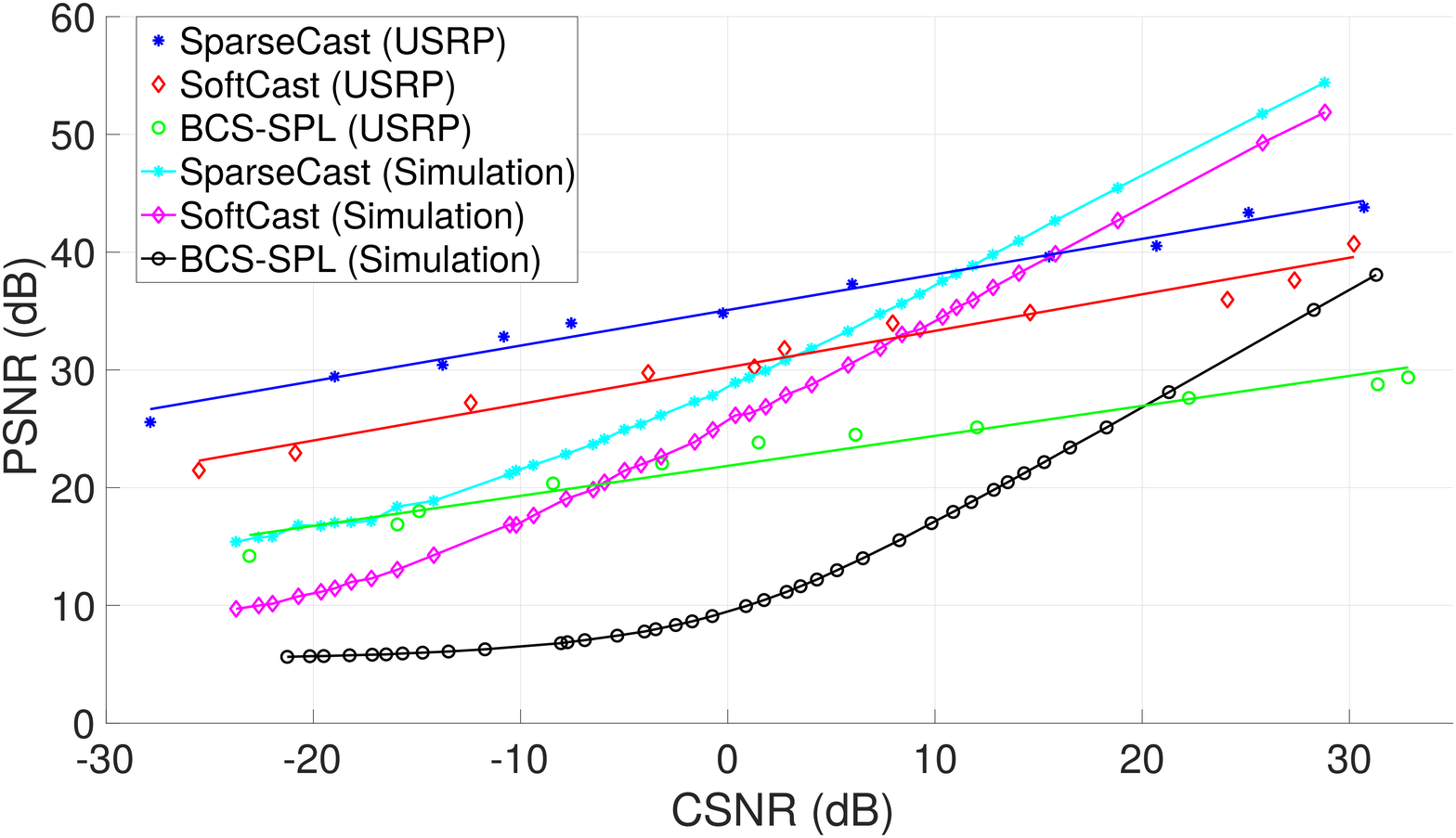}}
    \hfill
  \subfloat[Low bandwidth usage ($75,000$ channel symbols).]{%
  \label{subfig:75k_USRP_lenna}
       \includegraphics[width=\linewidth]{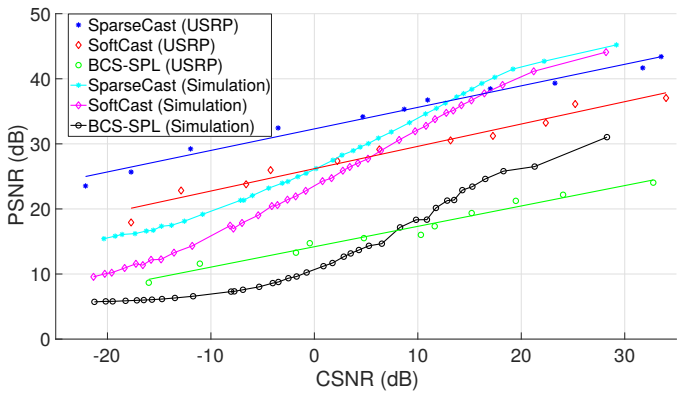} }
  \caption{USRP and simulation results. Block size for SparseCast is $16\times 16$ while it is $32\times 32$ for SoftCast and BCS-SPL. The lines drawn for USRP results are the linear regression lines for each respective algorithm. In (a), SparseCast was simulated with $\mu_j=3k_j$ and $\tau_K=0.1$, BCS-SPL with $\lambda=0.001$ and sampling ratio of $0.65$, SoftCast with block threshold of $7400$. In (b), SparseCast was simulated with $\mu_j=3k_j$ and $\tau_K=3.5$, BCS-SPL with $\lambda=0.001$ and sampling ratio of $0.4$, SoftCast with a block threshold of $28,000$.}
  \label{fig:USRP_CSNR_PSNR_lenna} 
  \vspace{-0.3cm}
\end{figure}

The meta-data size also differs across algorithms, with SparseCast requiring the greatest amount. In our simulations, SparseCast requires meta-data size of around $17,000$ bits with block size $16\times 16$. SoftCast requires $10,000-16,000$ bits depending on the block threshold, while BCS-SPL requires no meta-data, both with block size $32\times 32$. Different block sizes are used to ensure the meta-data size is similar. The increase compared to SoftCast is due to the need to transmit $\mu_j$'s and the way vectors $\mathbf{x}_j$ are generated. This is subject to change for different block sizes and parameters and should only have a small impact on the overhead of the algorithm. We emphasise that the meta-data size is parameter dependent and the results show that, under similar meta-data sizes, SparseCast is able to outperform SoftCast in terms of PSNR. It should be noted here that real numbers (i.e. mean and variance values) use the single precision floating point format (32 bits).

\begin{figure} 
    \centering
  \subfloat[Digital, $\mbox{PSNR}=25.00 dB$]{%
  \label{subfig:75k_digital_visual_lenna}
      \includegraphics[width=0.46\linewidth]{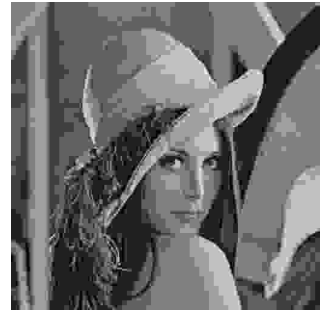} }
    \hfill
  \subfloat[SparseCast, $\mbox{PSNR}=31.05 dB$]{%
  \label{subfig:75k_proposed_visual_lenna}
        \includegraphics[width=0.46\linewidth]{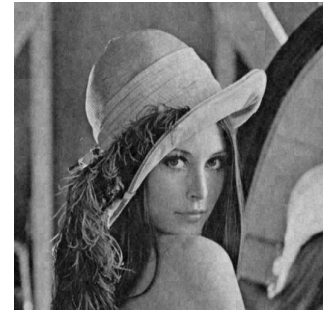}}
    \\
  \subfloat[SoftCast, $\mbox{PSNR}=27.98 dB$]{%
  \label{subfig:75k_softcast_visual_lenna}
        \includegraphics[width=0.46\linewidth]{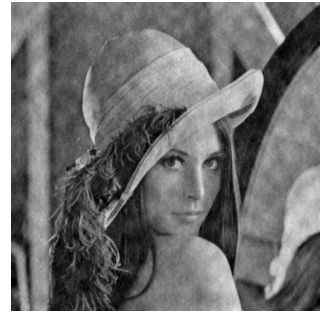}}
    \hfill
  \subfloat[BCS-SPL, $\mbox{PSNR}=14.13 dB$]{%
  \label{subfig:75k_spl_visual_lenna}
        \includegraphics[width=0.46\linewidth]{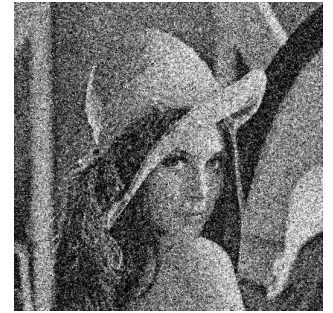}}  
  \caption{Simulation results for visual quality comparison of test image ``Lenna" under $\mbox{CSNR}=5 dB$ and channel symbol length $75,000$. For digital transmission JPEG compression is used with BPSK modulation and 1/2 rate convolutional coding.}
  \label{fig:75k_visual_comparison_lenna}
  \vspace{-0.3cm}
\end{figure}

Finally, the results obtained from the practical implementation confirm the observations made from simulations. As can be seen in Figure \ref{fig:USRP_CSNR_PSNR_lenna}, the same linear relationship between PSNR and CSNR can be observed. However, an important difference between the USRP and simulation results is the slope of the plots. Whereas the simulation results exhibit a slope of $1$, due to the AWGN assumption, the USRP results show slopes at about $1/2$. This is likely due to additional channel distortions caused by disturbances in the environment and transmitter/receiver oscillator misalignment. The latter is due to the free running oscillators in the USRP hardware, which can result in phase offset between the carrier frequencies.

The simulated results also tend to plateau at very low CSNR values as seen in Figure \ref{fig:USRP_CSNR_PSNR_lenna}. This effect is caused by the output pixel luminosity being limited to the range of $[0,255]$, limiting the maximum possible error. This cannot be seen in the USRP results as the PSNR never reached the level suggested by the simulation for the effects to be observed. To replicate this phenomenon with the USRP hardware would require a larger CSNR range which was not possible with the hardware model used due to antenna saturation and gain limitations. 

\section{Conclusions}
\label{sec:conclusion}

We proposed SparseCast, a novel hybrid digital-analog image transmission technique based on uncoded transmission of DCT coefficients. Sparsity in the frequency domain is exploited to improve the bandwidth usage, and a fixed set of measurement levels are used to reduce the amount of transmitted meta-data. We have used AMP to recover sparse vectors at the receiver to reduce decoder complexity. Shown by both simulation and experimental results, the combined use of MMSE and AMP for CS recovery overcomes the shortcomings of iterative CS recovery algorithms under non-sparse scenarios, while exploiting sparsity effectively. We have focused on the transmission of a single image here, but the proposed scheme can easily be used for video transmission, similarly to \cite{jakubczak_softcast:_2010}, \cite{li_video_2011} and \cite{schenkel_compressed_2010}. Non-linear mappings can also be used for bandwidth compression or expansion to better exploit the channel bandwidth as in \cite{Kim:DCC:08, Saleh:SPL:12} with an increased encoder and decoder complexity. 

\bibliographystyle{IEEEtran}
\bibliography{Final_ref}

\end{document}